\documentstyle[12pt]{article}
\input{epsfig.sty}
\newcommand{\beq}{\begin{eqnarray}}
\newcommand{\eeq}{\end{eqnarray}}
\begin{document}
\title{ $J/\Psi$, $\Psi(2S)$  Production in pp Collisions at E=510 GeV}
\author{Leonard S. Kisslinger$^{1}$\\
Department of Physics, Carnegie Mellon University, Pittsburgh PA 15213 USA.\\
Debasish Das$^{2,3}$\\
Saha Institute of Nuclear Physics,1/AF, Bidhan Nagar, Kolkata 700064, INDIA.}
\date{}
\maketitle
\noindent
PACS Indices:12.38.Aw,13.60.Le,14.40.Lb,14.40Nd
\begin{abstract}
  This brief report is an extension of studies of  $J/\Psi,\Psi(2S)$ 
production in pp collisions at the BNL with E=$\sqrt{s}$=200 GeV to  E=510 GeV 
at PHENIX.
\end{abstract}

\noindent
1) kissling$@$andrew.cmu.edu \hspace{1cm} 2)dev.deba$@$gmail.com; 
3) debasish.das@saha.ac.in

\section{Differential Rapidity Cross Sections for $J/\Psi,\Psi(2S)$ 
Production at E= 510 GeV}

  In the present work we use the theory described in detail in  Ref\cite{klm11}
with applications to BNL-RHIC, LHC, and Fermilab, based on the octet 
model\cite{cl96,bc96,fl96} for pp production of heavy quark states; and
used for studies of pp collisions for Upsilon production at forward
rapidities\cite{kd13}, and for heavy quark production at 7 TeV\cite{kd213}
and 8 TeV\cite{kd14}. This calculation is motivated by the report of 
preliminary data for $J/\Psi,\Psi(2S)$ production via pp collisions at 510 
GeV by the PHENIX Collaboration\cite{md14}.

  For helicity $\lambda$=0, the differential rapidity cross section is
given by\cite{klm11}
\beq
\label{dsig}
      \frac{d \sigma_{pp\rightarrow \Phi(\lambda=0)}}{dy} &=& 
     A_\Phi \frac{1}{x(y)} f_g(x(y),2m)f_g(a/x(y),2m) \frac{dx}{dy} \; ,
\eeq 
with with $a= 4m^2/s=3.46\times 10^{-5}$, $s=E^2$, $E=510$ GeV,
 $m=$ 1.5 GeV (for charm quark), and\cite{klm11}  $A_\Phi=\frac{5 \pi^3
 \alpha_s^2}{288 m^3 s} <O_8^\Phi(^1S_0)>$ =$3.1 \times 10^{-4}$ nb. 
$x(y)$ and $\frac{dx}{dy}$ are given by (there was a typo in the the numerator
of $\frac{d x(y)}{d y}$, with $(\exp{y}-\exp{(-y)})\rightarrow 
(\exp{y}+\exp{(-y)})$  in Ref\cite{klm11})
\beq
\label{1}
   x(y) &=& 0.5 \left[\frac{m}{510}(\exp{y}-\exp{(-y)})+\sqrt{(\frac{m}{510}
(\exp{y}-\exp{(-y)}))^2 +4a}\right] \nonumber \\
  \frac{d x(y)}{d y} &=&\frac{m}{1020}(\exp{y}+\exp{(-y)})\left[1. + 
\frac{\frac{m}{510}(\exp{y}-\exp{(-y)})}{\sqrt{(\frac{m}{510} 
(\exp{y}-\exp{(-y)}))^2 +4a}}\right] \; .
\eeq

The gluonic distribution function $f_g(x)$, for $\sqrt{s}$=E=510 
GeV\cite{klm11}, is
\beq
\label{2}
      f_g(x) & \simeq & 1334.21 - 67056.5 x + 887962.0 x^2 \; .
\eeq
 
\clearpage
From Eqs(\ref{dsig},\ref{1},\ref{2}) the differential rapidity cross sections
for $J/\Psi,\Psi(2S)$ production via 510 GeV p-p collisions with the standard
and mixed hybrid theories\cite{lsk09} are shown in the figure below.
\vspace{-1cm}

\begin{figure}[ht]
\begin{center}
\epsfig{file=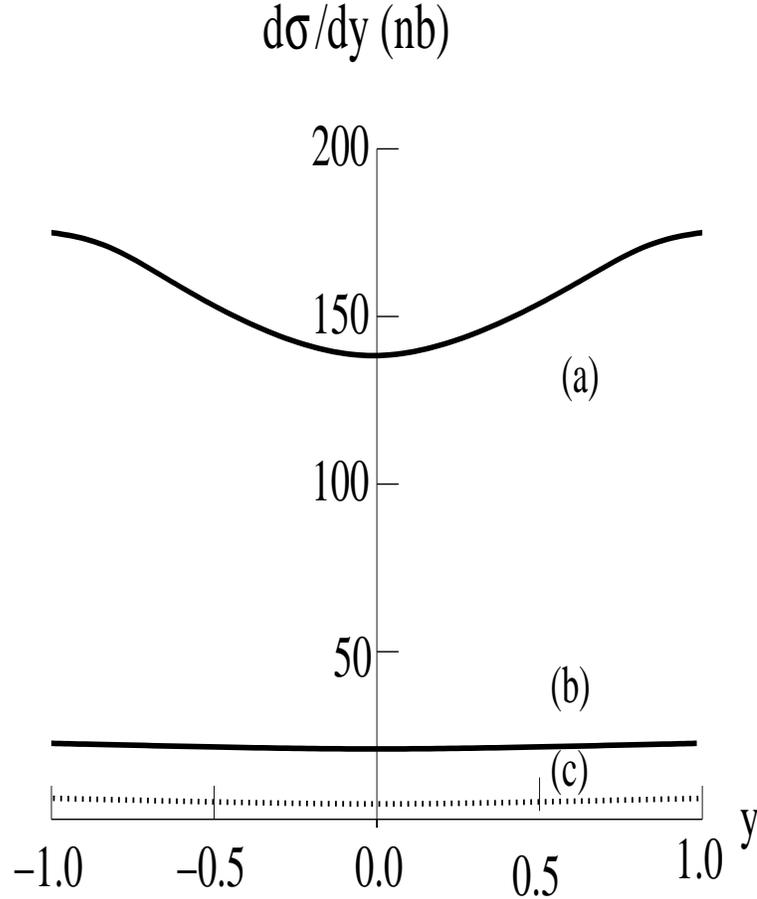,height=12cm,width=10cm}
\caption{$ d\sigma$/dy for E=510 GeV p-p collisions with $\lambda = 0$
producing (a)$J/\Psi$, (b) $\Psi(2S)$ with mixed hybrid theory, and (c)
 $\Psi(2S)$ with standard $c\bar{c}$ model.}
\end{center}
\end{figure}

\section{Conclusion}

  In anticipation of results from the PHENIX Collaboration\cite{md14}
(also see Ref\cite{wz14} for recent pp production of $J/\Psi$ and the
$(J/\Psi)/\Psi(2S)$ ratio at E=500 GeV)
we have calculated the differential cross sections for pp collisions
at 510 GeV for $J/\Psi$ production and $\Psi(2S)$ production both with
the standard $|c\bar{c}(2S)>$ model and the mixed heavy quark state hybrid 
theory\cite{lsk09}. As shown in the figure, the cross section for $\Psi(2S)$
production is 0.039 times that for $J/\Psi$ production in the standard
model, while in the mixed hybrid theory the factor is 0.122, approximately
a factor of $\pi$ larger, which has been found in experiments at 
200GeV\cite{klm11}. 

\Large{{\bf Acknowledgements}}

\vspace{5mm}
\normalsize 
Author D.D. acknowledges the facilities of Saha Institute of Nuclear Physics, 
Kolkata, India. Author L.S.K. acknowledges support from the P25 group at Los 
Alamos National laboratory. The authors thank Dr. Matthew Durham for helpful 
discussions.
 .

\end{document}